\documentclass[twocolumn,aps,superscriptaddress, showpacs]{revtex4}

\usepackage{graphicx}
\usepackage{latexsym}

\newcommand{\samp}{CaC$_6$ }

\newcommand{\pang}{$\textrm{\AA}^{-1}$}
\newcommand{\trans}{$_{ab}$ }
\newcommand{\refl}{$_c$ }
\newcommand{\Q}{$\vec{\textrm{Q}}$ }

\newcommand{\abp}{\mbox{\emph{ab}-plane} }

\begin{document}

%figure labels - width in gimp is 1020 35 pt arial
% font for x axis is 25 pt
%\preprint{APS/123-QED}

%\title{Inelastic X-ray Scattering Study of Phonons in Superconducting \samp}
\title{Phonons in Superconducting \samp Studied via Inelastic X-ray Scattering}

\author{M.~H.~Upton}
\affiliation{Condensed Matter Physics and Materials Science Department, Brookhaven National Laboratory, Upton, New York 11973, USA}
\author{A.~C.~Walters}
\affiliation{London Centre for Nanotechnology and Department of Physics and Astronomy, University College London, London WC1E 6BT, United Kingdom}
\affiliation{ISIS Facility, Rutherford Appleton Laboratory, Chilton, Didcot, United Kingdom}
\author{C.~A.~Howard}
\affiliation{London Centre for Nanotechnology and  Department of Physics and Astronomy, University College London, London WC1E 6BT, United Kingdom}
\author{K.~C.~Rahnejat}
\affiliation{London Centre for Nanotechnology and Department of Physics and Astronomy, University College London, London WC1E 6BT, United Kingdom}
\author{M.~Ellerby}
\affiliation{London Centre for Nanotechnology and Department of Physics and Astronomy, University College London, London WC1E 6BT, United Kingdom}
\author{J.~P.~Hill}
\affiliation{Condensed Matter Physics and Materials Science Department, Brookhaven National Laboratory, Upton, New York 11973, USA}
\author{D.~F.~McMorrow}
\affiliation{London Centre for Nanotechnology and Department of Physics and Astronomy, University College London, London WC1E 6BT, United Kingdom}
\affiliation{ISIS Facility, Rutherford Appleton Laboratory, Chilton, Didcot, United Kingdom}
\author{A.~Alatas}
\affiliation{Advanced Photon Source, Argonne National Laboratory, Argonne, Illinois 60439, USA}
\author{Bogdan~M.~Leu}
\affiliation{Advanced Photon Source, Argonne National Laboratory, Argonne, Illinois 60439, USA}
\author{Wei~Ku}
\affiliation{Condensed Matter Physics and Materials Science Department, Brookhaven National Laboratory, Upton, New York 11973, USA}

\date{\today}

\begin{abstract}
We investigate the dispersion and temperature dependence
 of a number of phonons in the recently discovered superconductor
CaC$_\textrm{6}$
utilizing inelastic x-ray scattering.  
Four [00L] and two \abp phonon modes are observed,
and measured at temperatures both above and below T$_\textrm{c}$.
In general, our measurements of phonon dispersions are in good agreement with
existing theoretical calculations of the phonon dispersion.  
This is significant in light of several discrepancies between
experimental measurements of phonon-derived quantities and 
theoretical calculations.
The present work suggests that the origin of these discrepancies lies in the
understanding of the electron-phonon coupling in this material, rather than
in the phonons themselves.
\end{abstract}


\maketitle

\samp has attracted a great deal of interest since the recent discovery of its 
superconductivity below 11.4 K \cite{Weller, EmeryStruct}.
The transition temperature is unprecedented in 
graphite intercalation compounds, where 
transition temperatures are, with one exception, an order of magnitude
lower.  (The second highest known
transition temperature, of YbC$_6$, is still nearly 50\% lower, 6.5 K.)
Despite an intriguing initial suggestion
of a novel pairing mechanism \cite{Csanyi}
much of the early experimental experimental evidence 
indicates that \samp is a conventional 
weakly-coupled superconductor \cite{Kim2, Lamura}. 
Still, the unusually high transition temperature
has prompted a great deal of speculation 
as to which phonons play a part in pairing and why the transition
temperature is so high
\cite{Calandra, Lamura, Kim1, Kim2, Mazin, Boeri}.

%In particular, Mazin has argued that the  high superconducting
%transition of \samp is due to participation of the intercalant
%electronic states at the Fermi  level \cite{Mazin}.
%These states couple
%strongly to the soft Ca phonon modes, which are then
%in turn responsible for
%the high superconductivity transition temperature.

Calandra and Mauri \cite{Calandra} and Kim et al.~\cite{Kim2}
have calculated the phonon spectra and the electron-phonon coupling, using
density functional theory (DFT).  Both predict electron-phonon coupling
with the Ca\refl (out-of-plane) and C\trans (in-plane) modes.
While these two  calculations of the phonon dispersions are similar, 
their predictions of the coupling differ, 
with Calandra and Mauri assigning additional coupling to the C\trans modes.

There are a number of discrepancies between the theoretically
calculated phonon spectra and electron-phonon
coupling and a variety of phonon-derived experimental quantities for this
system - 
as outlined in a recent paper by Mazin et al.~\cite{Mazin1}.
For example, recent measurements of the isotope effect, 
$\alpha(X)=-(d \log T_c/dM_X)$ (where 
$T_c$ is the superconducting transition temperature and
$M_X$ is the mass of atom $X$) have found 
$\alpha(\textrm{Ca}) = 0.53$ \cite{Hinks}, in disagreement with the calculated
value of  $\alpha(\textrm{Ca})=0.24$ \cite{Calandra}.
This suggests that the coupling to soft Ca modes is greater than 
predicted by calculation.
Furthermore,  
the measured temperature dependence of the specific heat
does not show the large coupling effects derived from
the calculated electron-phonon spectral function, 
$\alpha^2 F(\omega)$ \cite{Kim2, Mazin1}.  
(Also called the Eliashberg function,
$\alpha^2 F(\omega)$ is a measure of the electron-phonon 
coupling.  $F(\omega)$ is the phonon density of states \cite{Carbotte}.)
Thus, in contrast to the isotope effect measurements discussed above,
specific heat experiments suggest that the coupling to acoustic Ca modes 
is significantly weaker
than is predicted by theory.

Additionally, 
the upper critical field of \samp
has been shown to have a linear 
temperature dependence down to 1 K 
while calculations of the upper critical field,  based on 
 LDA calculations of $\alpha^2 F(\omega)$,
have predicted non-linear behavior
below 4 K \cite{Jobiliong, Kim1, Mazin1}.
Finally, DFT calculations of the phonon spectra
understate the increase
of T$_c$ under pressure by an order of magnitude 
\cite{CalandraAll, Kim2, Gauzzi}.

The measurements of the isotope effect and the specific heat appear
to lead to contradictory conclusions, since the isotope measurement implies 
larger-than-calculated coupling to soft Ca modes while
the specific heat measurements suggests less-than-calculated coupling
to soft Ca modes.
This apparent conflict
%contradiction 
rests on a comparison of experimental
results to theoretical predictions which yield 
$\lambda \sim 0.85$ \cite{Calandra, Kim2}.
However, a recent measurement suggests that the 
superconducting gap is 2.3 meV, 40\% larger than previously measured
\cite{Kurter}.  Such a large gap suggests that \samp is a 
strongly coupled superconductor.

It is clear that the theoretical understanding
of \samp is by no means complete, and in particular a number
of experimental observations of 
quantities that are either directly or indirectly
related to the phonon spectra are inconsistent with the present theoretical 
description.  
Given this situation 
it is clearly of great interest to measure the phonon spectra
and compare it to the existing theoretical predictions \cite{Calandra, Kim2}.

In this paper
we report measurements of the phonon dispersions in CaC$_6$,
utilizing inelastic x-ray scattering (IXS) to do so.
In particular, we have measured the 
dispersion, and temperature dependence of the dispersion,
for a number of the low energy modes, including soft 
Ca\trans modes and part of the
C$_{ab}$ mode, that are
believed to be important in the superconductivity of the \samp
\cite{Calandra, Kim1}.
In general, we find good agreement with the extant 
predictions for the phonon dispersion curves.  These results
apparently confirm the validity of the DFT approach used
in such calculations and suggest that the origin of the various discrepancies 
with experiment discussed above lie in the understanding of the
 electron-phonon coupling
not of the phonon modes themselves.

\begin{figure}
\includegraphics[width=3in]{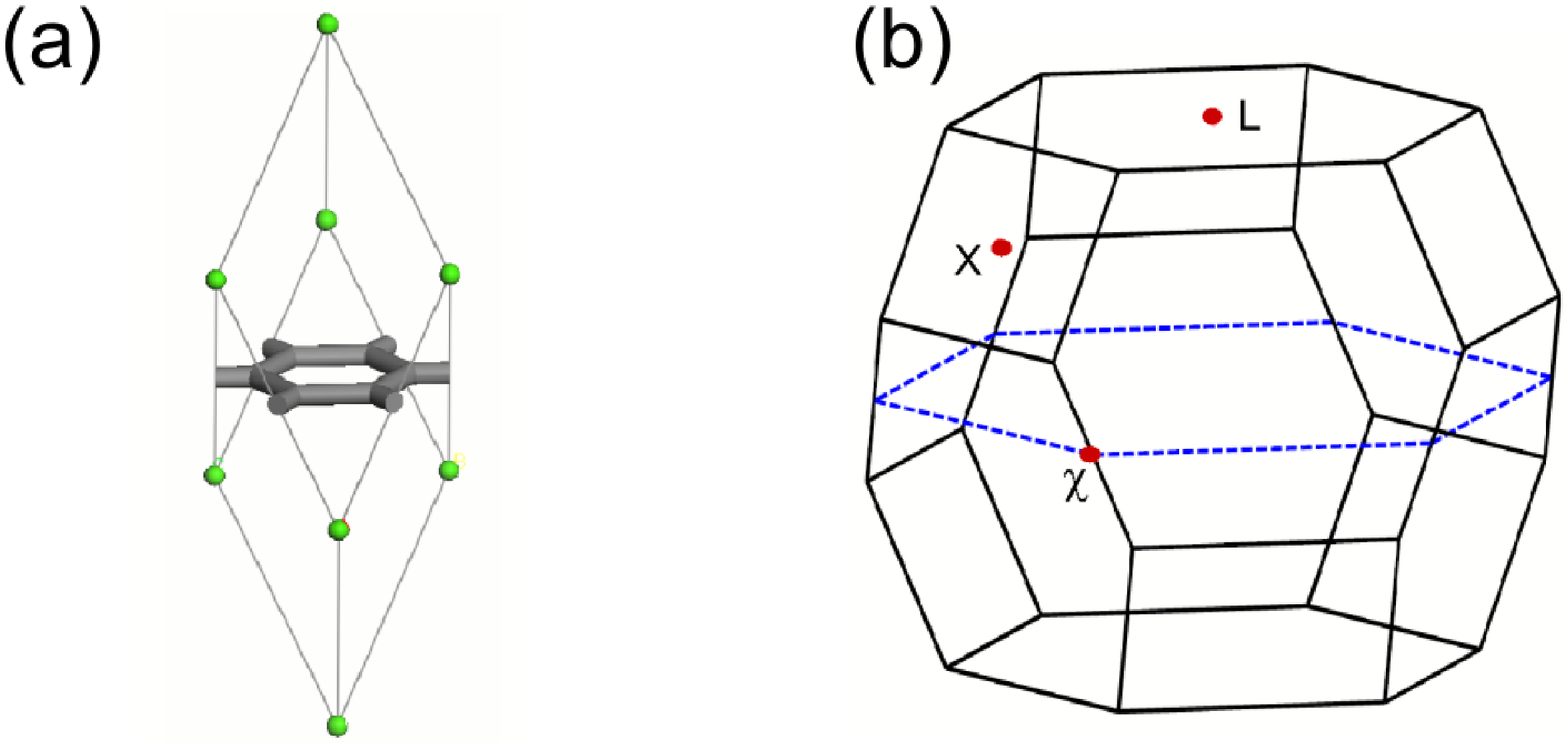}
\includegraphics[width=3in]{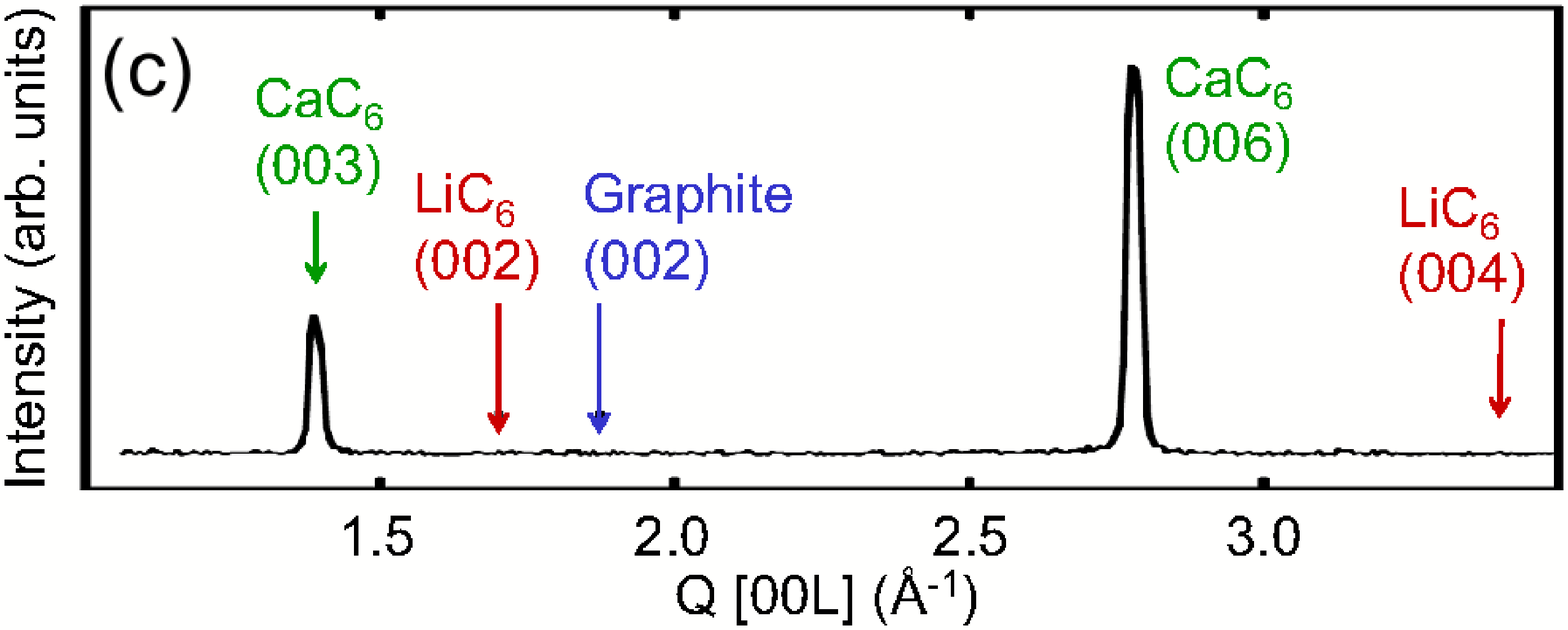}
\caption{\label{fig:Dif}
(color online) 
(a) The \samp crystal structure \cite{EmeryStruct}.  
Green circles represent Ca atoms and the grey ring is 
graphene \cite{RealSpaceAck}.
(b) The \samp Brillouin zone.
(c) An [00L] diffraction pattern taken after the inelastic experiment.  The
pattern exhibits sharp \samp (003) and (006) peaks.  
The locations of possible impurity peaks from  LiC$_6$ and Graphite 
are indicated.  Almost no scattering is observed at these locations.
}
\end{figure}

The samples were prepared via immersion of a Highly Orientated Pyrolytic 
Graphite (HOPG) platelet in a lithium/calcium alloy for 10 days,
as described in detail by Pruvost et al.~\cite{Pruvost}.
  The HOPG was ZYA grade purchased from GE Advanced 
Ceramics with an initial (non-intercalated) 
highly-aligned 
\emph{c}-axis mosaic of
 $\sim0.4^{\circ}$.  It is powder-like in the \emph{ab}-plane. The 
lithium and calcium
 were purchased from Sigma-Aldrich at 99.99\% purity.
  X-ray diffraction of the resulting shiny silver \samp platelet
is shown in Fig.~\ref{fig:Dif}c.
Diffraction showed very high sample purity with the 
intensity of the (002) Bragg peaks of the LiC$_6$ phase
and the graphite phase having less than 0.25\% the intensity of
the CaC$_6$ (006) Bragg peak. 
%(The  LiC$_6$ phase would indicate
%a lack of full intercalation 
%and the graphite phase would indicate a sample region with no intercalation.)

 The sample dimensions were 
$4 \times 4 \times 0.7 mm$ with a post-intercalation
 \emph{c}-axis mosaic of $\sim 3.5^\circ$.
The crystal structure and Brillouin zone of \samp
(space group $R\bar{3}m$, a=5.17 $\textrm{\AA}$,
$\alpha=49.55$) are shown in Figs.~\ref{fig:Dif}a and b \cite{EmeryStruct}.

%The crystal structure of \samp (space group $R\bar{3}m$, a=5.17 $\textrm{\AA}$,
%$\alpha=49.55$) is shown in Fig.~\ref{fig:Dif}a \cite{EmeryStruct}.  The
%Brillouin zone is shown in Fig.~\ref{fig:Dif}b.

After synthesis, the samples were 
mounted in a beryllium dome in an argon
or helium atmosphere.
The region around the beryllium dome was subsequently 
 pumped out to rough vacuum.  
The diffraction pattern of the sample was checked 
before, after and several times during the experiment. 
Both the out-of-plane diffraction patterns and the rocking curves of the 
(003) and (006) reflections were monitored 
during the experiments and
did not show
any signs of degradation \cite{DifNote}.
 A post-experiment
diffraction pattern is shown in Fig.~\ref{fig:Dif}c.

%%%%%%%%%%%%%%%%%%%%%%%%%%%%%%%%%%%%%%%%%%%%%%%%%%%%%%%%%%%%%%%%%%%%%%%
%%%%%%%%%%%%%%%%%%%%%%%%%%%%%%%%%%%%%%%%%%%%%%%%%%%%%%%%%%%%%%%%%%%%%%%

%%%%%%%%%%%%%%%%%%%%%%%%%%%%%%%%%%%%%%%%%%%%%%%%%%%%%%%%%%%%%%%%%%%%%%%
\begin{figure}
\includegraphics[width=3in]{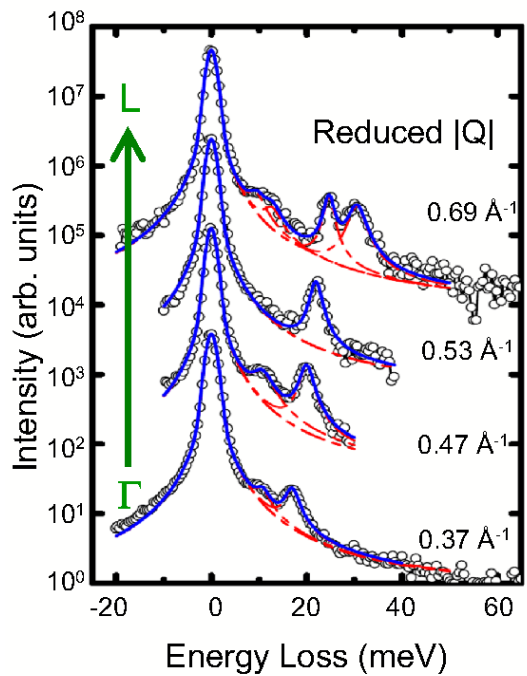}
\caption{\label{fig:spectra}
(color online) Inelastic spectra along $\Gamma-{\textrm L}$ 
from the 4th Brillouin
zone. 
As discussed in the text, the phonon cross-section 
 means that some of the phonon modes reported in Fig.~\ref{fig:BZ} 
do not appear in these data.
Data are labelled by momentum transfer in the reduced zone scheme.
The spectra are offset for clarity and are normalized to account
for the different efficiencies of the different detectors.
The y-axis is a $\log_{10}$ scale.
The blue (dark gray) long dashed lines are the total fit.
Each red (light gray) short-long dashed line
is a portion of the total fit, including the elastic line
and one phonon mode peak.}
\end{figure}
%C6Ca/April07/stackRefBZ5_5K.cpm
%%%%%%%%%%%%%%%%%%%%%%%%%%%%%%%%%%%%%%%%%%%%%%%%%%%%%%%%%%%%%%%%%%%%%%%

The IXS experiments were performed at sector 3
at the Advanced Photon Source of
Argonne National Laboratory \cite{Sinn1, Sinn2}.  Data
were collected using Si(18 6 3) analyzer reflections.  The 
instrument had an
overall resolution of 2.3 to 2.5 meV.  Photon flux was %in the range
 $1.8-2.4\times 10^9$ photons/second.  Four analyzer
crystals and four independent detectors 
allowed %were used to allow 
data collection 
at four momentum transfers simultaneously.
The momentum resolution was 0.072 \pang in 
the scattering plane
and 0.18 \pang perpendicular to it.
The beam size at the sample was $250 \times 300 \mu m$.
%For all work, 
The spectra were normalized by a beam intensity monitor
immediately before the sample.  
Spectra with \Q in the [00L] direction
 were measured in reflection geometry
while spectra with \Q in the \abp were measured in
transmission geometry.
Finally, we note that identical
results
were obtained from two samples, prepared at different times
and measured in separate runs.

%%%%%%%%%%%%%%%%%%%%%%%%%%%%%%%%%%%%%%%%%%%%%%%%%%%%%%%%%%%%%%%%%%%%%%%
\begin{figure}
\includegraphics[width=3in]{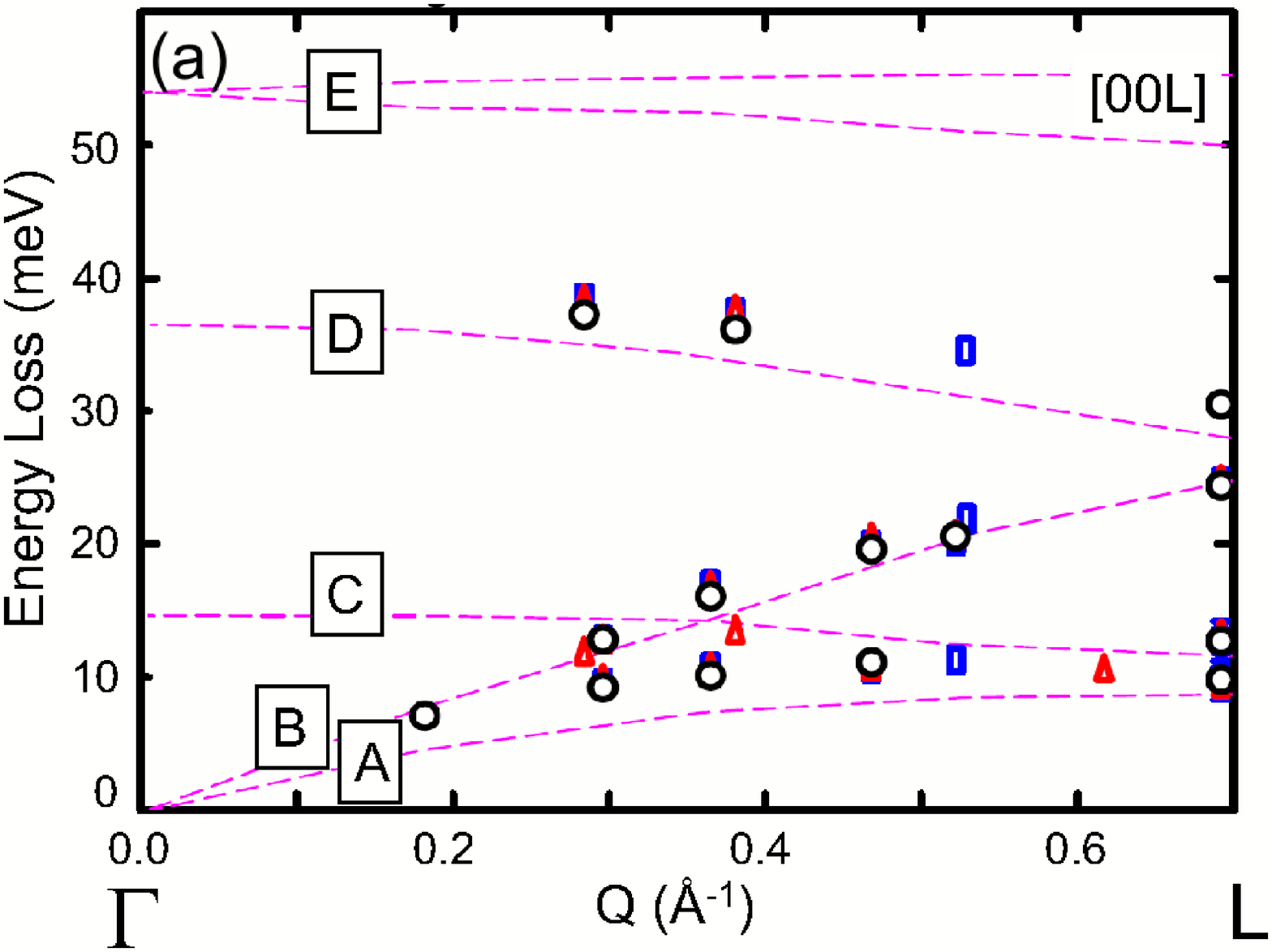}
\includegraphics[width=3in]{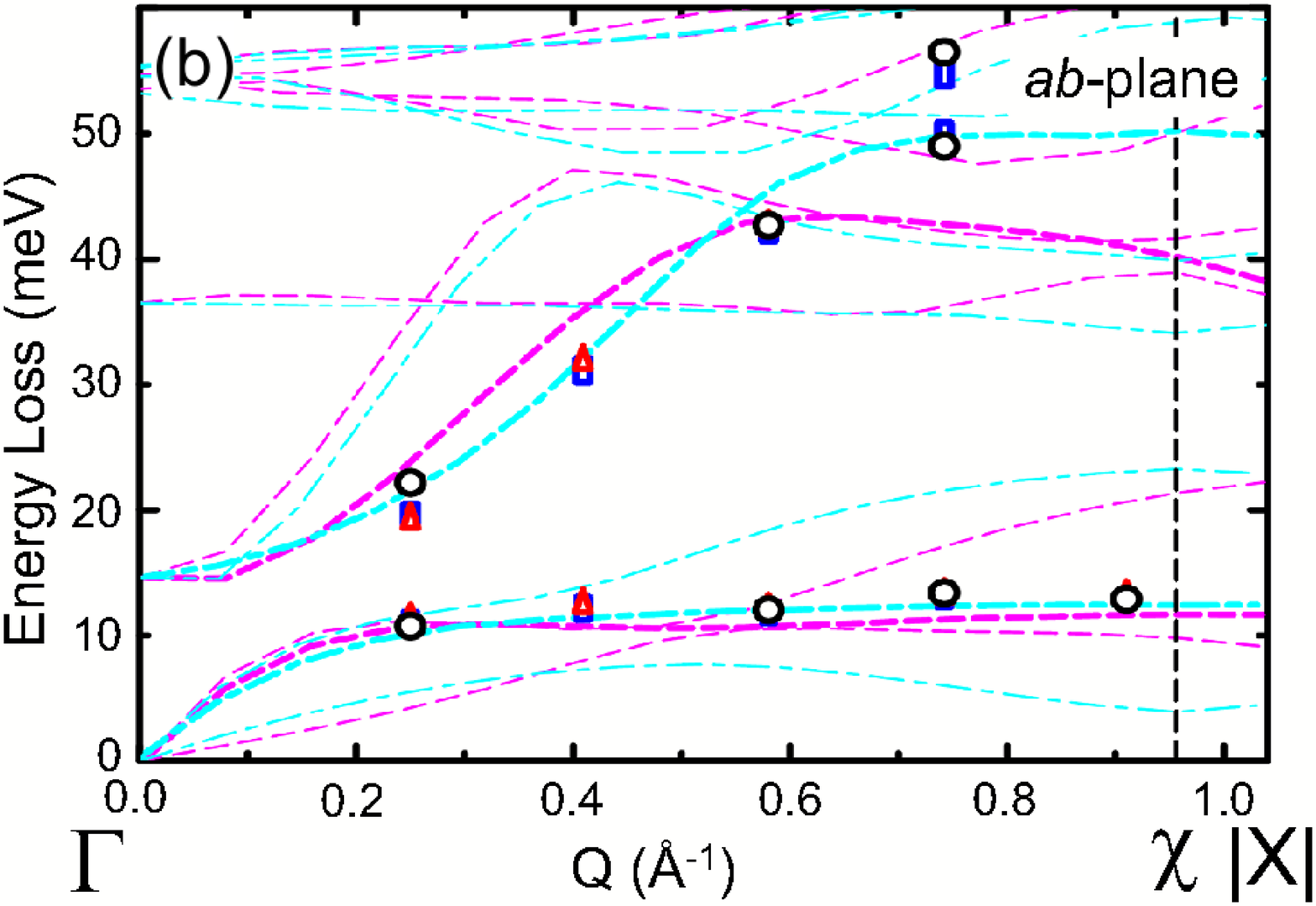}
\caption{\label{fig:BZ}
(color online) Phonon dispersion from spectra taken at three different
temperatures, 300 K in black circles, $\bigcirc$;
50 K in red triangles, $\bigtriangleup$, and 5 K shown 
in blue squares, $\Box$.  
(a) shows the [00L] spectra.  Calculated dispersions are shown as dashed
lines \cite{Calandra}.  
Different calculated modes are labeled by
A-E.
(b) shows the \abp spectra.
The distance to $\chi$
is indicated by a vertical black long dashed line.
The $\Gamma$-X calculated dispersions are shown as pink 
short-long dashed lines
and the $\Gamma$-$\chi$ dispersions are shown as blue dashed 
lines.
Note that $\Gamma$-X is not a strictly \emph{ab}-plane dispersion.
The theoretical curves most closely corresponding to the data shown
marked with thicker lines than other curves.
}
\end{figure}
%%%%%%%%%%%%%%%%%%%%%%%%%%%%%%%%%%%%%%%%%%%%%%%%%%%%%%%%%%%%%%%%%%%%%%%

A series of spectra taken at 5K with $\vec{\mathrm{Q}}$ 
in the [00L] direction 
and in the fourth Brillouin zone, with  $\vec{\mathrm{Q}}$ between 
$\frac{3}{2}\vec{\mathrm{G}}$ and $2\vec{\mathrm{G}}$ where  $\vec{\mathrm{G}}$
is one reciprocal lattice vector,
are shown in Fig.~\ref{fig:spectra}.
Typical phonon intensities are $<1$ count/second.  
Example fits are shown in the figure.  
The elastic line is fit with a psuedo-voigt while the phonon
peaks are fit with Lorentzians.
Near the L point zone boundary, the feature at 10 meV is 
seen to broaden and a better fit is obtained if the data are fit to two modes.

A summary plot showing the resulting peak positions
is shown in Fig.~\ref{fig:BZ}.
Note that not all of the phonon observations
reported in Fig.~\ref{fig:BZ}
are visible in Fig.~\ref{fig:spectra} 
because the cross-section of different phonons varies with
momentum transfer.
Specifically, the phonon intensity is proportional to 
$(\vec{\mathrm{Q}} \cdot \vec{\mathrm{\epsilon}})^2/\hbar \omega$,
 where $\vec{\mathrm{\epsilon}}$ is the atomic displacement, 
$\vec{\mathrm{Q}}$ is the momentum transfer and $\hbar \omega$ is the
energy of the phonon mode.
Some phonons shown in Fig.~\ref{fig:BZ}  
are visible only in higher Brillouin zones or
at longer counting times. 
The data in Fig.~\ref{fig:BZ}a are shown in a reduced zone 
scheme with  spectra from
different Brillouin zones folded back into the first Brillouin zone.
%Figure \ref{fig:BZ}b shows the \abp  modes.

We discuss first the $\Gamma -\textrm{L}$ dispersion (Fig.~\ref{fig:BZ}a).
Four phonon bands are observed in this direction.
The lowest energy band, labelled A, is almost
dispersionless over the measured region
and has an energy near 10 meV.
The second band, labelled B, disperses from \mbox{0 meV}
 to \mbox{25 meV}. 
A small portion of a third band, labelled C, can be seen near
L near 10 meV. The
final observed band, labelled D, disperses between \mbox{25 meV} and 
\mbox{40 meV}. 

It is difficult to assign an exact cross-section to a particular mode
because the data were collected in different geometries. 
The elastic intensity is a reasonable, but imperfect, 
measure of the amount of material in the scattering volume, 
and additional factors complicate
the comparison of data taken in the reflection ([00L]) and
 transmission (\emph{ab}-plane) geometries.
However, we can 
report the relative cross-sections of various modes at a particular momentum
transfer in the [00L] direction.
At \mbox{\Q=2.08 \pang}, which 
corresponds to a reduced \mbox{\Q =0.69\pang},  modes
C and D have roughly equal intensities of 0.5 counts/second while mode A 
has an intensity of roughly 0.1 counts/second at 5K.  These count rates
represent the raw data and are not scaled by the IXS polarization factor, 
$(\vec{\mathrm{Q}} \cdot \vec{\mathrm{\epsilon}})^2/\hbar \omega$.

Also shown in Fig.~\ref{fig:BZ}a are calculations of the phonon
bands \cite{Calandra}.  
The two calculations agree to within a few meV in the measured energy range
with the exception of the
dispersion of the lowest energy
optical band in the \emph{ab}-plane \cite{Calandra, Kim2}.
%While
%the two calculations are similar \cite{Calandra, Kim2}, 
%they differ slightly in
%the \emph{ab}-plane.  
The calculation of Calandra and Mauri \cite{Calandra}
agrees with the data somewhat better than the calculation
of Kim et al.~\cite{Kim2}, so it is shown here.
Moreover, Calandra and Mauri identify the phonon
mode associated with each calculated band \cite{Calandra}.  
By comparison of our experimental data with this calculation, we identify
mode A as a primarily Ca\trans 
oscillation with some C\refl
character, mode B as a mixed Ca\refl  and C\refl mode,
 and mode D as a mixed Ca\refl
and C\refl mode \cite{Calandra}.

In general, the agreement between theory and experiment is good,
though we note that the calculations of modes A and D 
understate the energy of the
measured band throughout the dispersion.
In addition, 
there are predicted modes which are largely or entirely 
%absent in our data; these are modes C and E.
absent in our data: modes C and E.
%We note that 
This apparent absence
does not necessarily mean that the mode does not occur.
% in the sample.  
In particular, mode C
is predicted to have
primarily Ca\trans character, which, 
because of the
phonon polarization factor in the IXS cross-section, 
means it should not be observed.
This mode also has a small amount of C\refl character, 
 in addition to the majority Ca\trans component.
%with a small
% amplitude of
%the C\refl oscillation relative the Ca\trans component.
A small C\refl amplitude may be why these measurements
observe the  C mode in only a small
portion of reciprocal space.
In fact, 
mode A is also predicted to have primarily Ca\trans character and a small
amount of C\refl character, consistent with this.
Near \mbox{\Q=1.24 \pang}
the measured intensity of this mode is 45 times larger in the \abp
than it is in the [00L] direction (after scaling by the elastic intensity).
Thus, the polarization factor also explains the asymmetry in the intensity
of this mode.
%/home/mhupton/C6Ca/April07/CPS
%factors such as Bragg structure

Despite measuring a wide range
around 50 meV we are not 
able to observe 50 meV C\refl phonons modes.
Spectra were counted for between 80 and 120 seconds per 0.25 meV step.
We believe that a 10 count peak would have been observable, which 
places an upper
bound on any such mode
of 0.1 counts/second at 300K, %room  temperature, or 
less than 10\% of the dispersing phonon band  at similar
momentum transfers.
%We note that 
A related C\refl phonon mode was also 
not observed in an IXS
measurement on %single crystal 
graphite \cite{Mohr}.
We speculate that %either 
the amplitude of these modes is very small or 
the lifetime is very short (resulting in a broad peak).  Either
of these effects, 
when 
combined with the
% relatively
 high predicted energy
 of mode E, 
would
result in
a very small IXS intensity.
%in figure 2 - 1 cps = 4mm, lowest point 1.25 cps, second 1.5cps, third 1cps

%dataCln/RT/CaC6_RT_11.1_A3_NoLiC6.dat around 0.75 cps second BZ 
%dataCln/RT/CaC6_RT_24.13_A3_NoLiC6.dat 2cps near 10meV, 4cps near 20 meV
%page 91,92 in lab book, 40 seconds per scan, at tth=15.5, 120 seconds tot
%RT, reflection

Next we turn to the \abp phonons (Fig.~\ref{fig:BZ}b).
The interpretation of the 
\abp phonons is somewhat complicated
because the data are not taken at a point in reciprocal space.  
Rather,
because the sample is not oriented in-plane, each spectra 
is taken at fixed $|\mathrm{\vec{Q}}|$,
which represents
an average over a circle in the \abp in reciprocal space
of radius $|\mathrm{\vec{Q}}|$, centered at $\Gamma$.
Therefore, 
 phonons which have a significant anisotropic
dispersion in the 
\abp
are not visible in these experiments
because this anisotropy will
translate into 
an  
energy  broadening of the measured mode and
consequently a reduction in the peak intensity.  
Of course,
if the \abp dispersion is isotropic (identical in all
\emph{ab}-plane 
 directions) the average is irrelevant and the mode 
may be observed.
%Additionally, 
Finally, 
all the data must be taken in the first Brillouin zone because
the powder structure of the \emph{ab}-plane
and the unequal distances  of different \emph{ab}-plane directions
in the Brillouin zone
 mean that phonon positions
cannot be folded back to the first Brillouin zone.
Despite these difficulties two phonon bands are observed.  
The 
%low energy,
flat  band near 10 meV and a highly dispersive  band which
we measure between 15 meV and 50 meV.
At \Q=0.58\pang and 5K, 
the two measured \abp phonon modes have roughly 
equal intensities of 0.1 counts/ second.  

Comparison with Calandra and Mauri 
(Fig.~\ref{fig:BZ}) identifies  the low energy band as a principally
Ca\trans mode and the dispersing band as a C\trans mode \cite{Calandra}.
This latter mode,
which is the \abp extension of 
mode C measured in the [00L] direction, discussed above,
 disperses to higher energy than 
predicted \cite{Kim2}.
A second highly dispersing C\trans mode is predicted by Calandra and
Mauri, but is not observed \cite{Calandra}.
This may be because the band is not isotropic.
In fact, the principal difference (below 60 meV) of the calculations
of Calandra and Mauri and Kim et al.~is that Calandra and Mauri predict
two isotropic C\trans bands dispersing from $\Gamma$ at 15 meV while
Kim et al.~predicts that only one of them is isotropic.

Calandra and Mauri \cite{Calandra} and Kim et al.~\cite{Kim2} both
predict a flat,
%relatively 
isotropic band near 38 meV.  This mode is not observed in 
experiments, %a fact
which cannot be due
to the \abp averaging effects discussed above. 
 Thus, 
its absence in the experimental data is explained by
%consistent  with
its assignment as a mixed Ca\refl and C\refl mode:
the IXS polarization factor would suppress any such mode in 
 this geometry.

Finally, the phonon dispersion is measured at three
 temperatures 5 K, which is below
the 11.4 K
transition temperature; 50 K and 300 K.
No differences in the phonon spectra at different temperatures are observed,
within the limits of our measurements.  Additionally, there 
are no significant changes in the widths.  Also, the intensity
of the peaks appears to follow the temperature
dependence predicted by Bose statistics, as expected.

In summary, we find good agreement between experiment and theory,
although calculation underestimates the 
energy of two modes in the [00L] direction.
This agreement has a number of implications for the general understanding
of the system.
One immediate consequence 
is that the predicted $F(\omega)$, the phonon density of states,
appears to be largely  correct 
\cite{Calandra,Kim2}.

Nevertheless, as discussed in the introduction, serious
disagreements exist between experiment and the theoretical
understanding  of the system.
The slight disparity between the calculated and measured $F(\omega)$
is probably not sufficient to reconcile these results.
The disagreements are largely a result of predictions 
made using the calculated  $\alpha^2 F(\omega)$ \cite{Calandra, Kim2}.
Thus,
 by confirming our understanding of the phonon dispersion, and hence
the calculated $F(\omega)$, the present work
raises serious questions about the calculation of the electron-phonon
coupling.  
We hope that this work inspires theoretical studies to reconcile the existing
discrepancies with experimental data.

%Additionally, this measurement lends weight to ancillary predictions made 
%by calculation, including predictions of $\alpha$, $\mu*$
%and $\lambda$, contrary to the measurements of Hinks et al.~\cite{Hinks}.

We acknowledge helpful discussions with T. Berlijn.
Work performed at BNL was supported by US DOE, Division of
Materials Science and Engineering, under contract No.~DE-AC02-98CH10886
and partially by DOE-CMSN.
Use of the Advanced Photon Source was supported by the US DOE,
Office of Science, %Office of 
Basic
Energy Sciences, under Contract
No.~DE-AC0Z-06CH11357.
Work performed at UCL was supported by the 
UK Engineering and Physical Science Research Council
and by a Wolfson Royal Society Research Merit Award.

\bibliography{CaC6Upton}

\begin{thebibliography}{22}
\bibitem[{Dif()}]{DifNote}
\bibinfo{note}{One of the first diffraction signs of degredation in air
  exposure is the increase in width of the (006) rocking curve relative to the
  (003) rocking curve.}
\bibitem[{Rea()}]{RealSpaceAck}
\bibinfo{note}{This figure was generated using a software program from
  Accelrys.}
\end{thebibliography}
\end{document}